\documentclass[twocolumn,showpacs,preprintnumbers]{revtex4}

\def\be{\begin{equation}}
\def\ee{\end{equation}}
\def\beq{\begin{eqnarray}}
\def\eeq{\end{eqnarray}}
\def\n{\nonumber}
\def\bay{\begin{array}}
\def\eay{\end{array}}

\begin{document}

\preprint{CIRI/01-swkg04}
\title{Axially Symmetric, Spatially Homothetic Spacetimes}

\author{Sanjay M. Wagh}
\affiliation{Central India Research
Institute, Post Box 606, Laxminagar, Nagpur 440 022, India\\
E-mail:ciri@vsnl.com}

\author{Keshlan S. Govinder}
\affiliation{School of Mathematical and Statistical
Sciences, University of Natal, Durban 4041, South Africa \\
E-mail: govinder@nu.ac.za\\ }
\bigskip

\date{January 7, 2002}

\begin{abstract}
We show that the existence of appropriate spatial homothetic
Killing vectors is directly related to the separability of the
metric functions for axially symmetric spacetimes. The density
profile for such spacetimes is (spatially) arbitrary and admits
{\it any} equation of state for the matter in the spacetime. When
used for studying axisymmetric gravitational collapse, such
solutions do not result in a locally naked singularity.
 \\

\centerline{Submitted to: Physical Review D - Rapid
Communications}
\end{abstract}

\pacs{04.20.-q, 04.20.Cv, 95.30.5f}%
\maketitle

\newpage
The Newtonian notion of self-similarity \cite{sedov} can be generalized in General
Relativity in different ways \cite{carrcoley}. It is important to distinguish
between these generalizations.

In general, the notion of any self-similarity or
scale-independence of the spacetime geometry requires the
spacetime to admit a homothetic Killing vector. A proper
homothetic Killing vector ${\bf X}$ satisfies \be {\cal
L}_{\scriptscriptstyle \bf X} g_{ab}\;=\;2\,\Phi\,g_{ab} \ee where
$\Phi$ is an arbitrary constant. This is also the broadest (Lie)
sense of the scale invariance of the spacetime leading to the
reduction of the Einstein field equations as partial differential
equations to ordinary differential equations. In particular, a
{\em spatially homothetic spacetime\/} admits an appropriate {\em
spatial\/} homothetic Killing vector.

On the other hand, the notion of {\it self-similarity} of physical fields requires
that the physical quantities transform according to their respective dimensions. The
self-similarity property of matter fields then requires restrictions on the
spacetime geometry. Such restrictions originate from the Einstein field equations
through the energy-momentum tensor of the matter.

In general, the self-similarity of geometry need not imply
self-similarity of physical fields. This leads to the question of
whether the matter fields exhibit the same symmetries as the
geometry - the so called ``symmetry'' inheritance problem.
Moreover, the self-similarity of physical fields need not, in
general, force the spacetime to admit a homothetic Killing vector.
This is the general problem of obtaining the constraints on the
metric from the self-similarity properties of matter fields - the
so called ``inverse'' symmetry inheritance problem. In particular,
we note that if the source of the spacetime is not perfect fluid,
the spacetime symmetries need not be obeyed \cite{coleytupper} by
the matter fields. Needless to say, it is important to distinguish
between all these cases.

We have recently shown \cite{cqg2} that the imposition of the
admission of a {\em spatial\/} homothetic Killing vector \be
\bar{X}_a=(0,f(r,t),0,0) \label{hkvnew} \ee on general spherically
symmetric spacetimes forces those spacetimes to be necessarily
separable in temporal and radial coordinates. The spacetime metric
that admits (\ref{hkvnew}) as a homothetic Killing vector is {\em
uniquely} the one reported in \cite{cqg1}, namely:
\begin{widetext} \be ds^2 = -\, y^2(r)\, dt^2 \;+\; \gamma^2\, B^2(t)\,(y')^2
\,dr^2 \;+\; y^2(r)\,Y^2(t)\, \left[\, d\theta^2 \,+\,
\sin^2{\theta}\,d\phi^2\,\right] \label{ssmetfinal} \ee
\end{widetext}
where an overhead prime denotes a derivative with respect to
$r$, $f(r,t)=y/(\gamma y')$ and $\gamma$ is a constant. (We absorb
a temporal function in $g_{tt}$ by a redefinition of $t$.) We
emphasize that the pure radial scale-independence requires the
spherically symmetric spacetime to admit a {\em spatial\/}
homothetic Killing vector (\ref{hkvnew}).

The Einstein tensor for (\ref{ssmetfinal}) has components
\begin{widetext} \beq G_{tt}&=& \frac{1}{Y^2}-\frac{1}{\gamma^2B^2} +
\frac{\dot{Y}^2}{Y} + 2\frac{\dot{B}\dot{Y}}{BY}
\\ G_{rr}&=&\gamma^2B^2
\left(\frac{y'}{y}\right)^2
\left[-\,2\frac{\ddot{Y}}{Y}-\frac{\dot{Y}^2}{Y}
+\frac{3}{\gamma^2B^2}+ \frac{1}{Y^2}\right]
\\G_{\theta\theta}&=&-\,Y\,\ddot{Y}-Y^2\frac{\ddot{B}}{B}
- Y\,\frac{\dot{Y}\dot{B}}{B}+\frac{Y^2}{\gamma^2B^2}
\\G_{\phi\phi}&=& \sin^2{\theta}\,G_{\theta\theta} \\
G_{tr}&=&2\frac{\dot{B}y'}{By}  \eeq \end{widetext}

Notice that $G_{tr}$ component of the Einstein tensor is
non-vanishing. Hence, matter in the spacetime could be {\em
imperfect\/} or {\em anisotropic\/} indicating that the
energy-momentum tensor could be \begin{widetext} \beq {}^{\rm
I}T_{ab}&=&(\,p\,+\,\rho\,)\,U_a\,U_b \;+\; p\, g_{ab}
\;+\;q_a\,U_b \;+\; q_b\,U_a \;-\;2\,\eta\,\sigma_{ab} \label{imperfect} \\
{}^{\rm A}T_{ab}&=&\rho \,U_a\,U_b \;+\; p_{||}\,n_a\,n_b \;+\;
p_{\bot}\,P_{ab} \label{anisotropic}  \eeq
\end{widetext} where $U^a$ is the matter 4-velocity, $q^a$ is the
heat-flux 4-vector relative to $U^a$, $\eta$ is the
shear-viscosity coefficient, $\sigma_{ab}$ is the shear tensor,
$n^a$ is a unit spacelike 4-vector orthogonal to $U^a$, $P_{ab}$
is the projection tensor onto the two-plane orthogonal to $U^a$
and $n^a$, $p_{||}$ denotes pressure parallel to and $p_{\bot}$
denotes pressure perpendicular to $n^a$. Also, $p$ is the
isotropic pressure and $\rho$ is the energy density.

Now, the Einstein field equations with imperfect matter of
vanishing shear viscosity yield for (\ref{ssmetfinal})
\begin{widetext} \beq \rho &=&
\frac{1}{y^2}\left(\frac{\dot{Y}^2}{Y^2} + 2 \frac{\dot{B}}{B}
\frac{\dot{Y}}{Y} + \frac{1}{Y^2} - \frac{1}{\gamma^2 B^2}\right)
\label{sepdens} \\2\,\frac{\ddot{Y}}{Y} \;+\;\frac{\ddot{B}}{B}
&=&\frac{2}{\gamma^2B^2} \;-\; \frac{y^2}{2}\,\left(
\,\rho\,+\,3\,p \right) \label{isopressure}  \\
q &=& - \frac{2 \dot{B}}{y^2 \gamma^2 y' B^3} \label{heatflux}
\eeq
\end{widetext}
where $q^a = (0, q, 0, 0)$ is the radial heat-flux vector. Therefore, the spacetime
of (\ref{ssmetfinal}) has the property that the temporal and radial metric functions
are determined independently of each other. The Einstein field equations do not
determine the radial function $y(r)$ \cite{cqg1}. On the other hand, the temporal
functions $B(t)$ and $R(t)$ are determined by the properties of matter generating
the spacetime such as its equation of state.

However, it is usual  \cite{psj} to enforce on the spherically
symmetric metric the form \be \tilde{X}_a = (T,R,0,0)
\label{hkvusual} \ee for the homothetic Killing vector. Then, all
the dimensionless quantities such as metric functions etc.\ are
functions only of the dimensionless self-similarity variable
$T/R$.

However, in \cite{prl2} we showed that, for spherically symmetric spacetimes, the
form (\ref{hkvusual}) of the homothetic Killing vector is too restrictive and
obscures important information about the properties of such spacetimes, for example,
the existence of naked singularities. This is understandable since the Killing
vector (\ref{hkvusual}) corresponds to the {\em simultaneous\/} scale-invariance of
the spacetime in $T$ and $R$ in the sense of Lie.

We emphasize that, for spherical symmetry, the appropriate form is
(\ref{hkvnew}) since it corresponds to only the radial
scale-invariance of the spacetime in the sense of Lie. However,
(\ref{hkvnew}) is equivalent to (\ref{hkvusual}) under the
transformation
\begin{widetext} \be R = l(t) \exp\left(\int f^{-1} d r\right)
\qquad T = k(t) \exp\left(\int f^{-1} d r\right)
\label{sstrans}\ee
\end{widetext}
Of course, there exists a relation between the temporal functions
$l(t)$ and $k(t)$ since the transformed metric can always be made
diagonal in $R$ and $T$ coordinates. It should be noted that
spacetimes admitting (\ref{hkvusual}) are included in
(\ref{ssmetfinal}) when the transformations (\ref{sstrans}) are
non-singular and, hence, are preserving the spatial character of
(\ref{hkvnew}). Such non-singular transformations preserve the
spatial scale-invariance property of gravity, in general.

Further, the spacetime of (\ref{ssmetfinal}) admits \cite{cqg2}
matter density that is an arbitrary function of the radial
coordinate $r$ since the field equations do not determine $y(r)$
and $\rho \propto 1/y^2$. This property of the spacetime, that of
admitting an arbitrary spatial density, then also relates to the
scale-invariance of the geometry in $r$. Hence, the homothetic
Killing vector (\ref{hkvnew}) indicates a fundamental property of
gravity that it has no radial scale for matter inhomogeneities.
Such spacetimes are then of fundamental importance to matters of
gravity.

Clearly, any imposition of the homothetic Killing vector
(\ref{hkvusual}) on a spherically symmetric spacetime is {\em
over-restrictive} and is not demanded by any {\em basic property
of gravitation}. It can then be shown \cite{sscollapse} that the
spacetime of (\ref{ssmetfinal}) does not lead to a naked
singularity when the density is initially non-singular. It follows
that the naked singularities can only arise when (\ref{sstrans})
are singular, that is when one of the basic properties of gravity,
the pure spatial scale-invariance, is violated.

We note that a perfect fluid spacetime cannot admit a non-trivial
homothetic Killing vector which is orthogonal to the fluid
4-velocity unless $p=\rho$ \cite{mcintosh}. The spacetime of
(\ref{ssmetfinal}) admits (\ref{hkvnew}) - a non-trivial, spatial
homothetic Killing vector orthogonal to the fluid 4-velocity. When
the temporal functions in (\ref{ssmetfinal}) are constants, the
equation of state for the matter, a perfect fluid now, is \be
p\,=\,\frac{1}{y^2}\left(\frac{4}{\gamma^2B^2}\,-\,\frac{2}{Y^2}
\right)\;+\;\rho \label{stateqstate} \ee  It is {\em uniquely\/}
$p=\rho$ since $B$, $Y$, $\gamma$ are arbitrary constants that can
be chosen suitably.

Encouraged by the example in spherical symmetry, in our current paper, we then
consider the implications of such a requirement of pure {\em spatial\/} homothety
for axially symmetric spacetimes. In axial symmetry we have two spatial variables
which can be expected to behave in a homothetic manner, viz. $r$ and $z$. In other
words, we expect the spacetime to admit arbitrary functions of $r$ and $z$
determining the matter characteristics of axially symmetric spacetimes.

Guided by these considerations, it would be natural to consider the existence of a
homothetic Killing vector of the form \be X_a=(0,f(r),g(z),0) \label{ashkv1} \ee for
the axisymmetric metric
\begin{widetext} \be ds^2 = -\,\bar{A}^2(t,r,z) dt^2+ \bar{C}^2(t,r,z)
dr^2 + \bar{D}^2(t,r,z) d z^2 +\bar{B}^2(t,r,z) d\phi^2 \label{asmet1} \ee
\end{widetext}
However, the form (\ref{ashkv1}) suggests a relationship between the $r$ and $z$
variables which is quite restrictive since it will correspond to {\em
simultaneous\/} scale-invariance of these variables in the sense of Lie. This is
contrary to our expectation that the matter characteristics in $r$ and in $z$ be
independently specifiable. We expect the spacetime to be determined independently in
these two spatial directions since gravity specifies no length-scale in either
variables and not just for their combination. For example, the density distribution
of a cylinder in $r$ and $z$ directions should, in general, be independently
specifiable and that too in any desirable manner since gravity provides no
length-scale for inhomogeneities in either variables.

Taking the above into account, we impose the existence of two
independent {\em spatial\/} homothetic Killing vectors of the form
\be {\bf H}_r = (0,f(r),0,0) \label{hkvr}\ee and \be {\bf H}_z =
(0,0,g(z),0) \label{hkvz} \ee on (\ref{asmet1}). This reduces the
metric (\ref{asmet1}) {\it uniquely\/} to
\begin{widetext} \be ds^2=\exp\left[\left(\int\frac{\Phi_1dr}{f(r)}
 + \int\frac{\Phi_2dz}{g(z)}\right)\right]\left[A^2(t) dt^2 + \frac{C^2(t)}{f^2(r)} dr^2
 +\frac{D^2(t)}{g^2(z)} dz^2 + B^2(t) d\phi^2\right] \ee
 \end{widetext} where the $\Phi$'s are the constant conformal factors.
Choosing $f(r)=\Phi_1 y(r)/y'$ and $g(z)=\Phi_2 Z(z)/\tilde{Z}$
where an overhead prime denotes differentiation with respect to
$r$ and an overhead tilde denotes differentiation with respect to
$z$, we then obtain
\begin{widetext} \be ds^2 = - Z^2y^2 dt^2 + \gamma_1^2 Z^2 C^2
(y')^2 dr^2 + \gamma_2^2 D^2 y^2 (\tilde{Z})^2dz^2 +
Z^2y^2B^2d\phi^2 \label{aximetfinal} \ee
\end{widetext}
where the $\gamma$'s are constants related to $\Phi$ s. We also absorb a temporal
function in $g_{tt}$ by a redefinition of the time coordinate.

The Einstein tensor for (\ref{aximetfinal}) has the components
\begin{widetext} \beq G_{tt}&=& -\frac{1}{\gamma_2^2D^2}-\frac{1}{\gamma_1^2C^2} +
\frac{\dot{C}\dot{D}}{CD} + \frac{\dot{B}\dot{D}}{BD}
+\frac{\dot{B}\dot{C}}{BC}\\ G_{rr}&=&\gamma_1^2C^2
\left(\frac{y'}{y}\right)
\left[-\frac{\ddot{D}}{D}-\frac{\ddot{B}}{B} -
\frac{\dot{B}\dot{D}}{BD}+\frac{3}{\gamma_1^2C^2}+
\frac{1}{\gamma_2^2D^2}\right] \\G_{zz}&=&\gamma_2^2D^2
\left(\frac{\tilde{Z}}{Z}\right)\left[-\frac{\ddot{C}}{C}-\frac{\ddot{B}}{B}
- \frac{\dot{B}\dot{C}}{BC}+\frac{3}{\gamma_2^2D^2}+
\frac{1}{\gamma_1^2C^2}\right] \\G_{\phi\phi}&=& B^2\left[
-\frac{\ddot{D}}{D}+\frac{1}{\gamma_2^2D^2}-\frac{\ddot{C}}{C}
-\frac{\dot{C}\dot{D}} {CD} +\frac{1}{\gamma_1^2C^2} \right] \\
G_{tr}&=&2\frac{\dot{C}y'}{Cy} \\
G_{tz}&=&2\frac{\dot{D}\tilde{Z}}{DZ}\\ \n \\
G_{rz}&=&2\frac{\tilde{Z}y'}{Xy} \eeq
\end{widetext} with an overhead dot denoting a time derivative.
It is clear from the above that the spacetime necessarily
possesses energy and momentum fluxes. The matter in the spacetime
is {\em imperfect}/{\em anisotropic}.

This is interesting in its own right. Any mass-particle of an axisymmetric body has
a Newtonian gravitational force directed along the line joining it to the origin.
This force, which is unbalanced during the collapse, has generally non-vanishing
components along $r$ and $z$ axes. Hence, a non-static axisymmetric spacetime of
(\ref{aximetfinal}) will necessarily possess appropriate energy-momentum fluxes!

It is important to note that the coordinates $(t, r, z, \phi)$ are
{\em not\/} co-moving. These coordinates are geared to the spatial
homothetic Killing vectors. The matter 4-velocity, in general,
will have all the four components, ie, $U^a\,=\,\left( U^t, U^r,
U^z, U^{\phi} \right)$.

In the case that $U^{\phi}=0$, the spacetime of
(\ref{aximetfinal}) describes any non-rotating, axisymmetric
matter configuration, in particular, a cigar configuration. In the
case that $U^{\phi}\neq 0$, the spacetime of metric
(\ref{aximetfinal}) describes {\em rotating\/} matter
configurations. In other words, it represents the ``internal''
Kerr spacetimes that are also axisymmetric in nature. (It is also
clear that non-static ``internal'' Kerr spacetimes cannot admit
any perfect fluid matter since axisymmetry requires the existence
of appropriate energy-momentum fluxes in such spacetimes as is
evident from the earlier discussion.)

The spacetime (\ref{aximetfinal}) has a singularity when either $C(t)=0$ or $D(t)=0$
for some $t$ or when $y(r)=0$ for some $r$ and/or $Z(z)=0$ for some $z$. Moreover,
from (\ref{aximetfinal}), the $r$ and $z$ null cone equations are \beq \frac{dt}{dr}
&=& \pm\,\gamma_1 \frac{y'}{y} C(t)
\\\frac{dt}{dz} &=& \pm\,\gamma_2 \frac{\tilde{Z}}{Z} D(t) \eeq
and these are non-singular for nowhere-vanishing functions $y(r)$
and $Z(z)$. Hence, there does not exist an out-going null tangent
at the spacetime singularity when $y(r) \neq 0$ and $Z(z) \neq 0$.
Hence, the singularities of these axisymmetric spacetimes are {\em
not\/} naked with these restrictions on the spatial functions.

We note that the nowhere-vanishing of $y(r)$ and $Z(z)$ means that the density is
initially non-singular. Moreover, it is also clear that the spacetime of
(\ref{aximetfinal}) will allow an arbitrary density profile in $r$ and $z$ since the
field equations do not determine these spatial functions. Further, it is also seen
that the spacetime of (\ref{aximetfinal}) admits {\em any\/} equation of state for
the matter in the spacetime and that the properties of matter in the spacetime
determine the temporal metric functions.

We also note that the energy-momentum tensor of the imperfect matter in the
spacetime can contain contributions from the presence of electromagnetic fields in
the matter. (That is why we listed only the Einstein tensor above.) The spacetime of
(\ref{aximetfinal}) can then be used to describe the process of accretion of matter
onto a rotating black hole. In this context, we note that the temporal behavior of
the spacetime is all that is determinable from the properties of matter including
those of the electromagnetic fields in the spacetime.

Moreover, a collapsing object could stabilize by the switching on
of some forces opposing gravity. Stable such objects correspond to
static spacetimes. Then, by considering temporal functions of the
spatially homothetic spacetimes, namely, (\ref{ssmetfinal}) and
(\ref{aximetfinal}), appearing in the energy fluxes to be
constants, we could obtain the spacetimes of stabilized objects
with corresponding symmetries. When the equation of state of
matter in a spatially homothetic, non-static spacetime changes to
that of the corresponding static spacetime during the collapse, we
obtain a stabilized object within these solutions. Note that we
need to use appropriate form of the energy-momentum tensor,
namely, (\ref{imperfect}) or (\ref{anisotropic}), to stabilize the
object.

Then, gravitational collapse may begin as dust but pressure must
build up, nucleosynthesis may commence to produce heat and may
result in a stabilized object like a star. The spatially
homothetic spacetimes reported here accommodate these features
because their temporal behavior is determined only by the
properties of matter such as its equation of state.

Further, one could argue that time ought to behave in a
self-similar manner as well and thus also impose a homothetic
Killing vector of the form \be {\bf H}_t=(h(t),0,0,0) \ee However,
this will impose a dynamic restriction on the spacetimes under
consideration. For example, it forces the spherically symmetric
spacetimes of (\ref{ssmetfinal}) to be shear-free. (See
\cite{cqg1,kill}.)

Moreover, in the case of axial symmetry, we also see that all the
dynamical and kinematical quantities are separable functions of
$t$, $r$ and $z$ under the imposition of the homothetic Killing
vectors (\ref{hkvr}) and (\ref{hkvz}). This is also the broadest
(Lie) sense in which the field equations as partial differential
equations reduce to ordinary differential equations and to their
separable form.

Therefore, the imposition of appropriate {\em pure} spatial
homothetic Killing vectors leads to spherically symmetric and
axisymmetric separable metric spacetimes that admit any equation
of state for {\em imperfect\/} and/or {\em anisotropic\/} matter.
The matter in these, generally non-static, spacetimes is imperfect
and/or anisotropic since General Relativity as a theory of gravity
seems to know, by virtue of its not providing any length-scale for
matter properties, that the inhomogeneous collapsing matter will
necessarily possess energy-momentum fluxes! The spacetimes
obtained here admit an {\em arbitrary\/} spatial distribution of
density. The spacetime singularity of such solutions is expected
to result from their temporal evolution when the spatial
distribution of density is initially non-singular. Such spacetime
singularities are not locally naked. This is consistent with the
Strong version of the Cosmic Censorship Hypothesis \cite{penrose}
which states that the singularities of gravitational collapse of
matter with regular, non-singular initial data should not be
visible to any observer, meaning that such singularities should
not be locally naked.

\section*{Acknowledgements}

We thank Pradeep S. Muktibodh for useful discussions. KSG thanks the University of
Natal and the National Research Foundation for ongoing support. He also thanks CIRI
and the Raman Research Institute for their kind hospitality during the course of
this work.

\end{document}